\documentclass[journal]{IEEEtran}
\usepackage{xr}
\usepackage{cite}
\usepackage{xspace}
\usepackage{amsmath,amssymb,amsfonts}
\usepackage{algorithm}
\usepackage{algpseudocode}

\usepackage{amsthm}
\usepackage{graphicx}
\usepackage{url}
\usepackage{xcolor}
\usepackage{mathtools}
\usepackage{makecell}
\usepackage{float}
\usepackage[font=small]{caption}
\usepackage{subcaption}
\usepackage{wrapfig}
\usepackage[switch]{lineno}
\usepackage{multirow}
\usepackage{multicol}
\usepackage{algpseudocode}
\usepackage{algorithm}

\usepackage[nopostdot,shortcuts,acronym]{glossaries}
\makeglossaries
\newacronym{caiso}{CAISO}{California Independent System Operator}
\newacronym[\glsshortpluralkey=DERs]{der}{DER}{distributed energy resource}
\newacronym[\glsshortpluralkey=ACE]{ace}{ACE}{area control error}
\newacronym{agc}{AGC}{automatic generation control}
\newacronym{nodes}{NODES}{Network Optimized Distributed Energy Systems}
\newacronym{bess}{BESS}{Battery Energy Storage System}
\newacronym{v2g}{V2G}{Vehicle-to-Grid}
\newacronym{ev}{EV}{electric vehicle}
\newacronym{ahu}{AHU}{air handling unit}
\newacronym{hvac}{HVAC}{heating ventilation and air conditioning}
\newacronym{ucsd}{UCSD}{University of California, San Diego}
\newacronym{pv}{PV}{photovoltaic}
\newacronym{iso}{ISO}{independent system operator}
\newacronym{pjm}{PJM}{Pennsylvania-New Jersey-Maryland Interconnection}
\newacronym{v1g}{V1G}{unidirectional V2G}
\newacronym{rc}{RC}{Ratio-Consensus}
\newacronym{pd}{PD}{primal-dual}
\newacronym{dana}{DANA}{Distributed  Approximate  Newton  Algorithm}
\newacronym{scada}{SCADA}{Supervisory Control  and  Data  Acquisition}
\newacronym{mse}{MSE}{mean-squared-error}
\newacronym{rmse}{RMSE}{root-mean-squared-error}
\newacronym{ros}{ROS}{Robotic Operating System}
\newacronym{arpae}{ARPA-e}{Advanced Research Projects Agency-Energy}

\DeclareMathOperator{\N}{\mathcal{N}}
\DeclareMathOperator{\I}{\mathcal{I}}
\DeclareMathOperator{\Lagr}{\mathcal{L}}
\newcommand{\Pref}{\ensuremath{\subscr{P}{ref}}}
\newcommand{\pu}{\ensuremath{\underline{p}}}
\newcommand{\po}{\ensuremath{\overline{p}}}
\newcommand{\zeros}{\ensuremath{\mathbf{0}}}

\newcommand{\revision}[1]{\textcolor{black}{#1}}

\newcommand{\real}{\ensuremath{\mathbb{R}}}

\newcommand{\subscr}[2]{#1_{\textup{#2}}}

\renewcommand{\epsilon}{\varepsilon}

\usepackage{psfrag}

\renewcommand{\baselinestretch}{.98}
\allowdisplaybreaks

\begin{document}

\title{Frequency Regulation with Heterogeneous Energy Resources: A Realization using Distributed Control}

\author{Tor Anderson$^*$ \quad Manasa Muralidharan$^*$ \quad Priyank Srivastava$^*$ \quad Hamed Valizadeh Haghi \\ Jorge Cort\'{e}s \quad Jan Kleissl \quad Sonia Mart\'{i}nez \quad Byron Washom

\thanks{Tor Anderson, Priyank Srivastava, Jorge Cort\'{e}s, and Sonia Mart\'{i}nez are with the Department of Mechanical and Aerospace Engineering, UC San Diego, CA, USA. Email: {\small {\tt \{tka001, psrivast, cortes, soniamd\}@ucsd.edu}}.
\newline\hspace*{.5em} Manasa Muralidharan, Jan Kleissl, and Hamed Valizadeh Haghi are with the Department of Mechanical and Aerospace Engineering and the Center for Energy Research, UC San Diego, CA, USA. Email: {\small \tt{m1murali@ucsd.edu, valizadeh@ieee.org, jkleissl@ucsd.edu}}.
\newline\hspace*{.5em} Byron Washom is with Strategic Energy Initiatives, UC San Diego, CA, USA. Email: {\small \tt{bwashom@ucsd.edu}}.
\newline\hspace*{.5em} This research was supported by the Advanced Research Projects Agency - Energy (ARPA-e) under the NODES program, Cooperative Agreement DE-AR0000695. \newline\hspace*{.5em} $^*$: These authors contributed to the paper equally. }  }

\maketitle

\begin{abstract}
This paper presents one of the first real-life demonstrations of coordinated and distributed resource control for secondary frequency response in a 
power distribution grid. A series of tests involved up to 69 heterogeneous active distributed energy resources consisting of air handling units, unidirectional and bidirectional electric vehicle charging stations, a battery energy storage system, and 107 passive distributed energy resources consisting of building loads and solar photovoltaic systems. The distributed control setup consists of a set of Raspberry Pi end-points exchanging messages via an ethernet switch. Actuation commands for the distributed energy resources are obtained by solving a power allocation problem at every regulation instant using distributed ratio-consensus, primal-dual, and Newton-like algorithms. The problem formulation minimizes the sum of distributed energy resource costs while tracking the aggregate setpoint provided by the system operator. We demonstrate accurate and fast real-time distributed computation of the optimization solution and effective tracking of the regulation signal over 40~min time horizons. An economic benefit analysis confirms eligibility to participate in an ancillary services market and demonstrates up to \revision{\$53k} of potential annual revenue for the selected population of distributed energy resources.
\end{abstract} 

\vspace*{-1ex}
\renewcommand{\IEEEiedlistdecl}{\IEEEsetlabelwidth{SONET}}
\printglossary[type=\acronymtype,title=List of Abbreviations,,nonumberlist,nogroupskip]
\renewcommand{\IEEEiedlistdecl}{\relax}

\section{Introduction}
Many recent efforts seek to integrate renewable energy resources with the power grid to reduce the carbon footprint. The high variability associated with wind and solar power can be balanced using \glspl{der} providing ancillary services such as frequency regulation. Consequently, there is a growing interest among market operators in DER aggregations with flexible generation and load capabilities to balance fluctuations in grid frequency and minimize \glspl{ace}. The fast ramping rate and minimal marginal standby cost put many DERs at an advantage against conventional generators and make them suitable for participation in the frequency regulation market.

The fast ramping rates reduce the required power capacity of DERs to only 10\% of an equivalent generator to balance a frequency drop within 30~s~\cite{ZAO-LMC-LA-MTM:19}.  However, most individual DERs have small capacities, typically on the order of kWs compared to 10~s of MW for conventional frequency control resources. Commanding the required thousands to millions of DERs to replace existing frequency regulation resources over a large balancing area entails aggregating DERs that are distributed at end points all over the grid on customer premises. The dynamic nature, large number, and distributed location of DERs requires coordination. This is in contrast to existing frequency regulation~\cite{MKM:14} implementation with conventional energy resources. For example, \ac{caiso} requires all generators to submit their bids once per regulation interval. Then, the setpoints are assigned centrally to all resources every 2-4~s without any consideration of operational costs~\cite{CAISO:12}. While distributed control has the potential to enable DER participation in the frequency regulation market (e.g.,~\cite{PS-CYC-JC:18-acc}), there is a general lack of large-scale testing to prove its effectiveness for widespread adoption by system operators. The 2017 National Renewable Energy Laboratory Workshop on Autonomous Energy Grids~\cite{NREL:17} concluded that ``A major limitation in developing 
new technologies for autonomous energy systems is that there are no large-scale test cases (...). These test cases serve a critical role in the development, validation, and dissemination of new algorithms''.

The results of this paper are the outcome of a project under the ARPA-e \ac{nodes} program\footnote{\url{https://arpa-e.energy.gov/arpa-e-programs/nodes}}, which postulates DER aggregations as virtual power plants that enable variable renewable penetrations of at least 50\%.
The vision of the NODES program was to employ state-of-the-art tools from control systems, computer science, and distributed systems to optimally respond to dynamic changes in the grid by leveraging DERs while maintaining customer quality of service. The NODES program required testing with at least 100 DERs at power. Here, we demonstrate the challenges and opportunities of testing on a heterogeneous fleet of DERs for eventual operationalization of optimal distributed control at frequency regulation time scales. 


\textit{Literature Review.} To the best of our knowledge,  real-world testing of frequency regulation by DERs has been limited. A \ac{v2g} \ac{ev}~\cite{WK-VU-KH-KK-SL-SB-DB-NP:08} and two \ac{bess}~\cite{MS-DS-AS-RT-RL-PCK:13} provided frequency regulation.
76 bitumen tanks were integrated with a simplified power system model to provide frequency regulation via a decentralized control algorithm in~\cite{MC-JW-SJG-CEL-NG-WWH-NJ:16}. In buildings, a decentralized control algorithm controlled lighting loads in a test room~\cite{JL-WZ-YL:17},  centralized frequency control was applied to an \ac{ahu}~\cite{YL-PB-SM-TM:15,EV-ECK-JM-GA-DSC:18}, an inverter and four household appliances~\cite{BL-SP-SA-MVS:18}, and four heaters in different rooms~\cite{LF-TTG-FAQ-AB-IL-CNJ:18}. A laboratory home with an EV and an AHU, and a number of simulated homes were considered for demand response in~\cite{KB-XJ-DV-WJ-DC-BS-JW-HS-ML:16} through an aggregator at a 10~s level. Technologies for widespread, but centrally controlled, cycling of air conditioners directly by utilities~cf.~\cite{SDGE} and aggregators are common place for peak shifting, but occur over time scales of minutes to hours. Industrial solutions enabling heterogeneous DERs to track power signals also exist, but they are either centralized, cf.~\cite{SC-PA:16} or require all-to-all communication~\cite{AT-SZ-SR:17}.

Our literature review exposes the following limitations: (i) centralized control or need for all-to-all communication~\cite{WK-VU-KH-KK-SL-SB-DB-NP:08,MS-DS-AS-RT-RL-PCK:13,YL-PB-SM-TM:15,EV-ECK-JM-GA-DSC:18,BL-SP-SA-MVS:18,LF-TTG-FAQ-AB-IL-CNJ:18,KB-XJ-DV-WJ-DC-BS-JW-HS-ML:16,SDGE,SC-PA:16,AT-SZ-SR:17}, which does not scale to millions of DERs; (ii) small numbers of DERs~\cite{WK-VU-KH-KK-SL-SB-DB-NP:08,MS-DS-AS-RT-RL-PCK:13,YL-PB-SM-TM:15,EV-ECK-JM-GA-DSC:18,BL-SP-SA-MVS:18,LF-TTG-FAQ-AB-IL-CNJ:18,KB-XJ-DV-WJ-DC-BS-JW-HS-ML:16}; (iii) lack of diversity in DERs~\cite{WK-VU-KH-KK-SL-SB-DB-NP:08,MS-DS-AS-RT-RL-PCK:13,MC-JW-SJG-CEL-NG-WWH-NJ:16,JL-WZ-YL:17,YL-PB-SM-TM:15,EV-ECK-JM-GA-DSC:18,LF-TTG-FAQ-AB-IL-CNJ:18}, with associated differences in tracking time scales and accuracy. No trial has been reported that demonstrated generalizability to a real scenario with (i) scalable distributed control and a (ii) large number of (iii) heterogeneous DERs.

\textit{Statement of Contributions.} 
To advance the field of real-world testing of DERs for frequency control, we conduct a series of tests using a group of up to 69 active and 107 passive heterogeneous DERs on the University of California, San Diego (UCSD) microgrid~\cite{BW-JD-DW-JK-NB-WT-CR:13}. To the best of the authors' knowledge, this is the first work to consider such a large, diverse portfolio of real physical DERs for secondary frequency response. As such, the major contributions of this work are:
\begin{itemize}
    \item A detailed account of the testbed, including the DER actuation and sampling interfaces, the distributed optimization setup, and communication framework.
    \item A description of techniques to work around technical barriers, provision of lessons learned, and suggestions for future improvement. 
    \item Evaluation of the performance of both the cyber and physical layers, including an evaluation of eligibility requirements for and the economic benefit of participating in the ancillary services market.
\end{itemize}

\textit{Paper Overview.} 
Frequency regulation is simulated on the UCSD microgrid using real controllable DERs (Section \ref{ders}) to follow the \ac{pjm} RegD signal~\cite{PJM-signal:19} interpolated from 0.5Hz to 1Hz (Sections \ref{regulation-signal}). The DER setpoint tracking is formulated as a power allocation problem at every regulation instant (Section \ref{optimization-statements}), and uses three types of provably convergent distributed algorithms from~\cite{ADDG-CNH-NHV:12,AC-JC:16-allerton,AC-BG-JC:17-sicon,TA-CYC-SM:18-auto} to solve the optimization problem; see Appendix~\ref{sec:appendix}. 
Setpoints are computed distributively on multiple Raspberry Pi's communicating via ethernet switches (Section \ref{computing-setup}). The setpoints are implemented on up to 176 DERs at power using dedicated command interfaces via TCP/IP communication (Section \ref{actuation-interface}), the DER power outputs monitored (Section \ref{power-measurements}), and their tracking performance evaluated (Section \ref{error-metrics}). Results (Section~\ref{sec:results}) for the various test scenarios described in Section~\ref{sec:test-scenario} show that the test system tracks the signal with reasonable error despite delays in response and inaccurate tracking behavior of some groups of DERs, and qualifies for  participation in the PJM ancillary services market 
.

\section{Problem Setting}
This paper validates real-world DER controllability for participation in secondary frequency regulation through demonstration tests implemented on a real distribution grid. The tests showcase the ability of aggregated DERs to function as a single market entity that responds to frequency regulation requests from the \ac{iso} by optimally coordinating DERs. The goal is to monitor and actuate a set of real controllable DERs to collectively track a typical \ac{agc} signal issued by the ISO. 

Three different distributed coordination schemes optimize the normalized contribution of each DER to the cumulative active power signal. Unlike simulated models, the use of real power hardware exposes implementation challenges associated with measurement noise, sampling errors, data communication problems, and DER response. To that end, precise load tracking is pursued at timescales that differ by DER type consistent with individual DER responsiveness and communication latencies, yet meet frequency regulation requirements in aggregation. 

The 69 kV substation and 12 kV radial distribution system owned by UCSD to operate the 5~km$^2$ campus was the chosen demonstration testbed. It has diverse energy resources with real-time monitoring and control capabilities, allowing for active load tracking. This includes over 3~MW of solar \ac{pv} systems, 2.5~MW/5~MWh of BESS, building \ac{hvac} systems in 14 million square feet of occupied space, and over 200 \ac{v1g}~\revision{\cite{CAISO:14}} and V2G EV chargers. The demonstration tests used a representative population of up to 176 such heterogeneous DERs to investigate tracking behavior of specific DER types as well as their cooperative tracking abilities.  While the available DER capacity at UCSD far exceeds the minimum requirements for an ancillary service provider set by most ISOs (typically $\sim$ 1~MW), logistical considerations and controller capabilities dictated the choice of a DER population size with less aggregate power capacity (up to 184~kW) for this demonstration. Since this magnitude of power is insufficient to measurably impact the actual grid frequency, we chose to simulate frequency regulation by following a frequency regulation signal.

\section{Test Elements}\label{sec:elements}
Here, we elaborate on the different elements of the validation tests. 
These include the optimization formulation employed to compute  DER setpoints (Section~\ref{optimization-statements}), the reference AGC signal (Section~\ref{regulation-signal}) and types of DERs used to track it (Section~\ref{ders}), the computing platform (Section \ref{computing-setup}), the actuation  (Section~\ref{actuation-interface}) and monitoring interfaces (Section~\ref{power-measurements}),  the performance metrics used to assess the cyber and physical layers, and eligibility for market participation (Section \ref{error-metrics}).

\subsection{Optimization Formulation}\label{optimization-statements}
The optimization model for AGC signal tracking using DERs can be mathematically stated as a separable resource allocation problem subject to box constraints as follows: 

\begin{equation}\label{eq:opt}
\begin{aligned}
\underset{p\in\real^n}{\text{min}} \
&f(p) = \sum_{i=1}^n f_i(p_i), \\
\text{s.t.} \ &\sum_{i=1}^n p_i = \Pref, \\
&p_i\in [\pu_i, \po_i], \quad \forall i\in\N = \{1,\dots,n\}.
\end{aligned}
\end{equation}

The agents $i\in\N$ each have local ownership of a decision variable $p_i\in\real$, representing an active power generation or consumption quantity (setpoint), a local convex cost function $f_i$, and local box constraints $[\pu,\po ]$, representing active power capacity limits. $\Pref$ is a given active power reference value determined by the ISO and transmitted to a subset of the agents as problem data, see e.g.~\cite{CAISO:18}. $\Pref$ is a signal that changes over time, so a new instance of~\eqref{eq:opt} is solved in \revision{real-time} 1~s intervals corresponding to these changes. \revision{Note that with just 1~s difference between the instances, the box constraints might also change due to the limited ramp rates of DERs. In this work we consider them constant and assume~\eqref{eq:opt} is feasible.}

For the validation tests, we used two types of cost functions: constant and quadratic. Constant functions were used for the Ratio-Consensus (RC) solver~\revision{\cite{ADDG-CNH-NHV:12}}, which turns the optimization into a feasibility problem. Quadratic functions were used for the primal-dual based (PD)~\revision{\cite{AC-JC:16-allerton,AC-BG-JC:17-sicon}} and Distributed Approximate Newton Algorithm (DANA)~\revision{\cite{TA-CYC-SM:18-auto}} methods. \revision{In short, RC prescribes dynamics which seek to achieve \emph{consensus} on a \emph{ratio} of operating capacity with respect to $\pu_i,\po_i$ so that the agents achieve $\sum_i p_i = \Pref$. PD and DANA  each are Lagrangian-based dynamics; in particular, PD is gradient-based (``first-order") and DANA is Newton-based (``second-order"). See Appendix~\ref{sec:appendix} for more technical detail on these algorithms}. 
The quadratic functions were artificially chosen to produce satisfactorily diverse and representative solutions \revision{to~\eqref{eq:opt}} for each DER population. \revision{Costs associated with a physical or economic metric (e.g. deviation from a building setpoint for AHUs, user-specified charging demands for V1G and V2Gs, and resistive losses in a BESS) are of great interest, but are far from trivial to model and thus not the focus of this study.} We split the total time period of the signal, $\Pref$ into three equal segments, and implemented RC, PD, and DANA in that order. Box constraints $[\pu_i,\po_i ]$ \revision{are given in Table~\ref{table:devicerating} and} were centered at zero for simplicity; \revision{for example, an AHU $i$ with 2~kW capacity has $[\pu_i,\po_i] = [-1,1]$, while a V2G $j$ with $\pm$5~kW capacity has $[\pu_j,\po_j] = [-5,5]$. }

\subsection{Regulation Signal 
}\label{regulation-signal}
\begin{figure}[hbt!]
\centering 
\includegraphics[width=\linewidth]{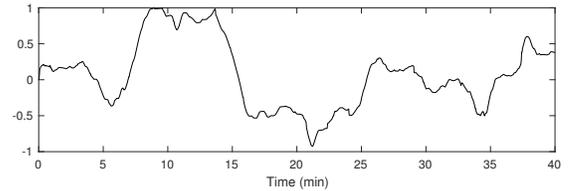}
\caption{\revision{Normalized PJM RegD signal.}}\label{fig:regd}
\vspace*{-2ex}
\end{figure}

The 40~min RegD signal published by PJM~\cite{PJM-signal:19} served as the reference AGC signal for the validation tests, and was used to obtain the value for $\Pref$ in~\eqref{eq:opt}. The normalized RegD signal, contained in $[-1,1]$ \revision{(see Figure~\ref{fig:regd})}, was interpolated from 0.5~Hz to 1~Hz. The signal was then treated by subtracting the normalized contributions of building loads and PV systems, cf. Section~\ref{ders}. Finally, the normalized signal was scaled by a factor proportional to the total DER capacity $\sum_i (\po_i - \pu_i)$ before sending to the optimization solvers. More precisely,
\begin{equation}\label{eq:norm-sig}
    \Pref = \beta \frac{\sum_i (\po_i - \pu_i)}{\| P_\text{RegD} + P_\text{PV} - P_\text{b} \|_{\infty} }\left(P_\text{RegD} + P_\text{PV} - P_\text{b}\right),
\end{equation}
where $P_\text{RegD}$ refers to the normalized RegD signal data, $P_\text{PV}$ and $P_\text{b}$ respectively refer to the normalized PV generation and building load data obtained from the UCSD ION server as described in Section \ref{power-measurements}, and $0 < \beta < 1$ is an arbitrary scaling constant.
%
\revision{Note that this results in a different target signal $\Pref$ for the different test scenarios considered in Section~\ref{sec:test-scenario} due to the different power ratings of the DERs (cf. Section~\ref{ders}) used across the tests.}
For most test scenarios, $\beta = 0.75$ to prevent extreme set points that would require all DERs to operate at either $\po_i$ or $\pu_i$ simultaneously, which may be infeasible in some time steps due to slower signal update times, see Table~\ref{table:devices}. Each $P$ in~\eqref{eq:norm-sig} is a vector with 2401 elements corresponding to each 1~s time step's instance of~\eqref{eq:opt} over the 40~min time horizon. 

\subsection{DERs}\label{ders}
The reference AGC signal was to be collectively tracked using DERs consisting of HVAC AHUs, BESS, V1G and V2G EVs, PV systems, and whole-building loads. Since PV systems and (non-AHU) building loads were not controllable, they participated in the test as passive DERs. Consequently, the active DERs were commanded to track a modified target signal derived by subtracting the net active power output of passive DERs from the reference AGC signal and applying appropriate scaling (cf. Section \ref{regulation-signal}). \revision{Table~\ref{table:devicerating}} lists the typical net power capacity $\po_i - \pu_i$ of the different active DER types.


\begin{table}[tbh]
\centering
\caption{\revision{Typical power rating of active DER types}}\label{table:devicerating}

\begin{tabular}{|c|c|c|c|c|}
\hline
\textbf{DER Type}  & \textbf{AHU} & \textbf{V1G EV} & \textbf{V2G EV} & \textbf{BESS} \\ \hline
\makecell{\textbf{Typical power} \\ \textbf{rating per DER type}} & 2~kW & \makecell{3.3~kW  (Tests 0 \& 1), \\ 4.9~kW  (Test 2)} & \revision{$\pm$} 5~kW & \revision{$\pm$} 3~kW \\ \hline
\end{tabular}
\end{table}

The contribution of each active DER to the target signal was defined with respect to a baseline power, around which $[\pu_i,\po_i ]$ was centered, to enable tracking of both positive and negative ramps in the target signal. For DERs like V2G EVs and BESS, which were capable of power adjustments in both directions, the baseline was 0~kW. The baseline for V1G EVs was defined to be halfway between their allowed minimum and maximum charging rates, where the former was restricted by the SAE J1772 charging standard to 1.6~kW. Similarly, the baseline for AHUs was defined to be half of their power draw when on. Further, since AHUs were limited to binary on-off operational states, the continuous and arbitrarily precise AHU setpoints obtained by solving \eqref{eq:opt} were rounded to the closest discrete setpoint obtained from a combination of on-off states before actuation.

AHU control was restricted, by UCSD Facilities Management, to specifying only DER setpoints and duration of actuation; since building automation controllers could not be modified, model-based designs were impossible. This was to avoid malfunctioning or disruptions to real physical infrastructure in the networked building management system that also controls lighting, security, and fire protection systems. 

\begin{figure*}[tbh]
\centering 
\includegraphics[width=\linewidth]{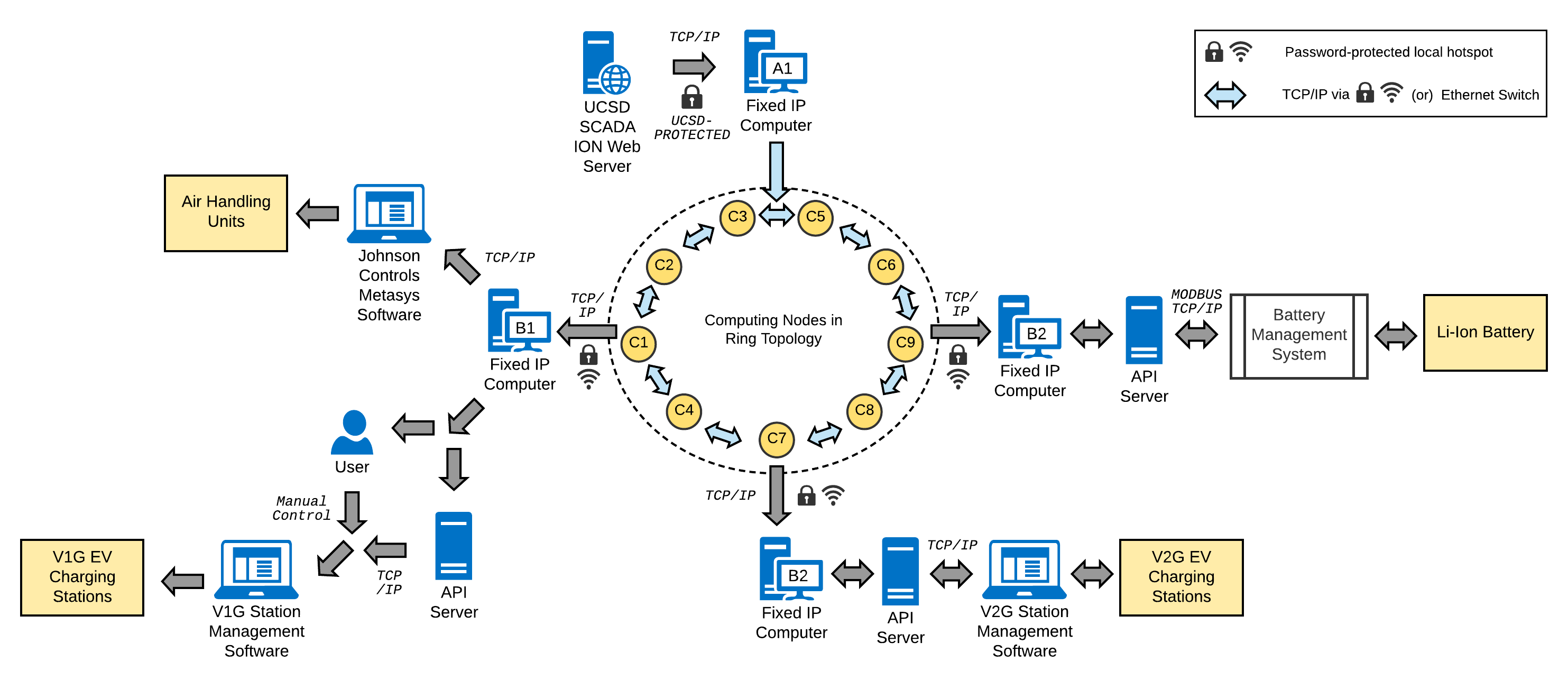}
\caption{Communication architecture for computation and actuation of control policies.}\label{fig:network-diagram}
\end{figure*}

\subsection{Computing Setup}\label{computing-setup}
The DER active power setpoints \revision{were computed for the entire 40-min test horizon prior to any device actuation using a set of 9 Linux-based nodes. The nodes C1-C9 }communicate with each other over an undirected ring topology, cf. Fig.~\ref{fig:network-diagram}. As one of the sparsest
network topologies, where message passing occurs only between a small number of neighbors, the ring topology presents a challenging scenario for distributed control. 
Since there were more active DERs than computing nodes, the 9 nodes were mapped subjectively to the 69 active DERs such that nodes C1-C2 computed the actuation setpoints for the AHUs, C3 for V1G EVs, C4-C8 for V2G EVs and C9 for the BESS. 

\revision{The computing steps are summarized in Algorithm \ref{alg:computing}.} Each computing node generated actuation commands as CSV files containing the power setpoints for their respective group of DERs at a uniform update rate of 1~Hz. Preliminary testing revealed different response times across DER types, with AHUs and V1G EVs exhibiting slower response than other active DER types. DERs with response times greater than 1~s were subject to a stair-step control signal with a signal update time consistent with DER responsiveness and constant setpoints during intermediate time steps. Table~\ref{table:devices} lists the signal update times for the different DER types.

\floatname{algorithm}{\color{black}Algorithm}
\begin{algorithm}
	\caption{\revision{Computing process}}
	\label{alg:computing}
	\color{black}
		\begin{algorithmic}[1]
		\Require {Map $f: C_i \rightarrow$ {DER-type}}
		\State Initialize time of last solution update $t_{\texttt{sol-update}_i} = 0$, initial setpoints for DERs mapped to computing node $C_i$ as $P_{f(C_i)},  \forall i\in\{1,\dots,9\}$
		\For {$k=0,\dots,2400$}
		\For {$i=1,\dots,9$} 
        \If{$k - t_{\texttt{sol-update}_i} == t_{\texttt{signal-update}_i}$}
		\State {Solve~\eqref{eq:opt} to update $P_{f(C_i)}(k)$}
		\State $t_{\texttt{sol-update}_i} = k$
		\EndIf
		\State $P_{f(C_i)}(k) \gets P_{f(C_i)}(t_{\texttt{sol-update}_i})$
		\If{$\operatorname{mod}(k,60) == 0$}
		\State Send $P_{f(C_i)}(k)$ to DER type, $f(C_i)$
		\EndIf
		\EndFor
		\EndFor 
	\end{algorithmic}
\end{algorithm}

\subsection{Actuation Interfaces and Communication Framework}\label{actuation-interface}
The actuation commands were issued using fixed IP computers through dedicated interfaces that varied by DER type as depicted in Fig.~\ref{fig:network-diagram}. The setpoints for AHUs were issued through a custom Visual Basic program that interfaced with the Johnson Control Metasys building automation software. The power rate of the BESS was set via API-based communication with a dedicated computer that controlled the battery inverter. The V1G and V2G EVs charging rates were adjusted through proprietary smart EV charging platforms of the charging station operators. EVs using ChargePoint\textsuperscript{\tiny\textregistered} V1G stations were manually controlled via the load shedding feature of ChargePoint’s station management software. The actuation of EVs using PowerFlex\textsuperscript{\tiny\textregistered} V1G chargers and Nuvve\textsuperscript{\tiny\textregistered} V2G chargers was automated and commands were issued via API-based communication.

\subsection{Power Measurements}\label{power-measurements}
The active power of all DERs was metered at a 1~Hz frequency. The power outputs of \revision{individual} PV systems and building loads were obtained prior to the test from their respective ION meters by logging data from the UCSD ION \ac{scada} system \revision{and aggregated to obtain the total power output of all PVs and building loads}. A moving average filter with a 20~s time horizon was used to remove noise from the \revision{aggregate} measured data for these passive DERs. V2G EVs and BESS power data were acquired using the same interfaces that were used for their actuation, which logged data from dedicated power meters.

Since neither AHUs nor the ChargePoint V1G EVs had dedicated meters, they were monitored via their respective building ION meters by subtracting a baseline building load from the building meter power output. Assuming constant baseline building load, any change in the meter outputs can be attributed to the actuation of AHUs and V1G EVs. This assumption is justifiable considering the tests were conducted at 0400 PT to 0600 PT on a weekend, when building occupancy was likely zero and building load remained largely unchanged. Noise in the ION meter outputs observed as frequent 15~-~30~kW spikes in the measured data for AHUs (Fig.~\ref{fig:trial0trial1}) and ChargePoint V1G EVs was treated by removing outliers and passing the resulting signal through a 4~s horizon moving average filter. Here, outliers refer to points that change in excess of 50\% of the mean of the   40~min signal in a 1~s interval.

\subsection{Performance Metrics}\label{error-metrics}
The performance of the distributed implementation (cyber-layer) was measured by the normalized \ac{mse} between the distributed and true (i.e. exact) centralized optimization solutions. The true solutions were computed for each instance of~\eqref{eq:opt} using a centralized CVX solver in MATLAB~\cite{website:cvx}. The MSE was normalized by dividing by the mean of the squares of the true solutions. 

The tracking performance of the DERs was evaluated through (i) the \ac{rmse} in tracking 
\begin{equation}\label{eq:rmse-calc}
    \text{RMSE} = \sqrt{\frac{\sum_{t=1}^T (P_t^{\text{prov}}-P_t^{\text{tar}})^2}{\sum_{t=1}^T (P_t^{\text{tar}})^2}},
\end{equation}
where $P_t^{\text{prov}}$ is the total power that was provided (measured), and $P_t^{\text{tar}}$ is the target (commanded) regulation power at time step $t\in\{1,\dots,T=2401\}$; and (ii) the tracking delay, computed as the time shift of the measured signal which yields the lowest RMSE between the commanded and measured signals. \revision{
The sum of the delays due to local computation and communication between the computing nodes is capped by the algorithm computation time, and would be less than 1~s.
Therefore, these delays are not explicitly considered in the tracking delay calculation, and the computed tracking delay only includes the device response times and measurement delays.}

The PJM Performance Score~$S$ following~\cite[Section 4.5.6]{PJM:20} was computed as a test for eligibility to participate in the ancillary services market, and is given by the mean of a Correlation Score~$S_c$, Delay Score~$S_d$, and Precision Score~$S_p$:
\begin{align*}
            S_c &= \frac{1}{T-1}\sum_{t=1}^T \frac{(P_t^{\text{prov}} - \mu^{\text{prov}})(P_t^{\text{tar}} - \mu^{\text{tar}})}{\sigma^{\text{prov}}\sigma^{\text{tar}}}, \\
        S_d &= \bigg\lvert \frac{\delta - 5 \text{ min}}{5 \text{ min}} \bigg\rvert, \quad
        S_p = 1 - \frac{1}{T} \sum_{t=1}^T \bigg\lvert \frac{P_t^{\text{prov}} - P_t^{\text{tar}}}{\mu^{\text{tar}}} \bigg\rvert, \\
        S &= 1/3(S_c + S_d + S_p),
        \end{align*}
where $P_t^{\text{prov}}$ and $P_t^{\text{tar}}$ are as in~\eqref{eq:rmse-calc},  $\mu^{\text{prov}}, \mu^{\text{tar}}$ and $\sigma^{\text{prov}}, \sigma^{\text{tar}}$ denote their respective means and standard deviations, and $\delta$ is the corresponding maximum delay in DER response for when $S_c$ was maximized. A performance score of at least 0.75 is required for participating in the PJM ancillary services market.

\section{\revision{Test Scenarios} 
}\label{sec:test-scenario}
\revision{In this section, we describe the test scenarios carried out on the UCSD microgrid 
elaborating on the challenges we faced and the differences across the tests, summarized by type of DER in Table~\ref{table:devices}.} 

\subsection{Commonalities}
A series of three tests were conducted on December 12, 2018 (Test~0), April 14, 2019 (Test~1) and December 17, 2019 (Test~2). All three tests involved a 40~min preparatory run followed by a 40~min final test. \revision{Table~\ref{table:devices} lists the 
type of DERs across the tests. All tests were carried out during non-operational hours (between 0400 PT and 0540 PT) to avoid potential disruptions to building occupants with the exception of V1G EVs in Test 2, which were tested at the start of the work day (0900 - 1010 PT) to maximize fleet EV availability (cf. Section~\ref{test-2}).} Day-time PV output data from February 24, 2019 was used as a proxy for an actual daytime PV signal.

\begin{table}[tbh]
\centering
\caption{\revision{Characteristics of each test by DER type.}}\label{table:devices}
\begin{tabular}{|>{\color{black}}c|>{\color{black}}c|>{\color{black}}c|>{\color{black}}c|>{\color{black}}c|}
\hline
\textbf{DER Type}           & \textbf{AHU}   & \textbf{V1G EV}                                                                     & \textbf{V2G EV} & \textbf{BESS} \\ \hline
\textbf{\# DERs - Test 0} & 7              & 4 
& 5 
& 1             \\ \hline
\textbf{\# DERS - Test 1} & 34             & 29 
& 5 
& 1             \\ \hline
\textbf{\# DERs - Test 2} & 34             & 17 
& 6 
& 1             \\ \hline
\textbf{Signal updates}  & 1 m          & \begin{tabular}[c]{@{}c@{}}5 m \\ (Tests 0 \& 1),\\ 1 m \\ (Test 2)\end{tabular}      & 1 s         & 20 s       \\ \hline
\textbf{DER Actuation} &
\begin{tabular}[c]{@{}c@{}}Synchronous\\ (Tests 0 \& 1),\\ Two-stage: Stage 1 \\(Test 2)\end{tabular} &
 \multicolumn{3}{>{\color{black}}c|}{\begin{tabular}[c]{@{}c@{}}Synchronous \\ (Tests 0 \& 1),\\ Two-stage: Stage 2 \\ (Test 2)\end{tabular}} \\
 \hline
\textbf{Operation Mode}     & Automatic      &
\begin{tabular}[c]{@{}c@{}}Manual \\(Tests 0 \& 1),\\ Automatic \\(Test 2)\end{tabular} & \multicolumn{2}{>{\color{black}}c|}{Automatic}  \\ \hline
\textbf{Time of test}       & 0400 - 0500 PT & \multicolumn{3}{>{\color{black}}c|}{\begin{tabular}[c]{@{}c@{}}0400 - 0500 PT (Tests 0 \& 1),\\ 0900 - 1010 PT (Test 2)\end{tabular}} \\ \hline
\textbf{Computing setup} &
  \multicolumn{4}{>{\color{black}}c|}{\begin{tabular}[c]{@{}c@{}}Semi-centralized using ROS (Tests 0 \& 1),\\ Fully distributed using Raspberry Pi (Test 2)\end{tabular}} \\ \hline
\end{tabular}
\end{table}

\subsection{Test~0}\label{test-0}
Test~0 was a preliminary calibration that \revision{was used 
to examine the response times and tracking behavior of every DER type and detect issues related to communication and actuation. 
\subsubsection{DERs} Test~0 used only a representative sample of 17 DERs. The V1G and V2G population was composed of UCSD fleet EVs plugged in at ChargePoint and Nuvve charging stations, respectively.
\subsubsection{Computing Setup} 9 laptops running a Robotic Operating System (ROS) communicated via local Wi-Fi hotspot to implement the distributed coordination algorithms and compute the DER setpoints.}
\subsubsection{\revision{Actuation}}
\revision{All DERs were actuated synchronously.}

\subsection{Test~1}\label{test-1}
Test~1 was identical to Test~0 
except in the number of DERs utilized.
\subsubsection{DERs} \revision{Test~1 used a larger population of 69 active DERs and 107 passive DERs.} 
\subsubsection{Computing Setup} 
\revision{ The same semi-centralized ROS-based computing setup as in Test~0 was used in Test~1.} Given that the available power capacity of fast-responding DERs such as V2G and BESS was smaller than slow-responding DERs, the steep ramping demands of the target signal were met by upscaling the power of the fast responding DERs in solving for the contribution of individual DERs. Another option would have been to reduce the number of slow responding DERs, but the funding agency stipulated prioritizing the number and types of heterogeneous DERs over accuracy in signal tracking. A real DER aggregator would instead require a more balanced capacity of slow and fast DERs to ensure feasibility of tracking these ramp features.
\subsubsection{\revision{Actuation}}
\revision{All DERs 
were actuated synchronously. Since the ChargePoint V1G EVs in Test~1 were operated via manual input of DER setpoints (an interface to their API had not been developed yet), to avoid overloading the (human) operators, they were grouped into three groups and actuated in a staggered fashion such that each of the three groups maintained a signal update time of 5~min but were commanded 1~min apart from each other.}

\subsection{Test~2}\label{test-2}
Test~2 also used the entire population of DERs but substituted the cumbersome V1G population with more capable V1G chargers and used a new distributed computing setup and method of actuation based on lessons learned from Test~1. 

\subsubsection{DERs} The V1G EVs used in Test~1 performed poorly owing to an unreliable actuation-interface that experienced seemingly random stalling and lacked automated control capabilities. Therefore, 17 PowerFlex V1G charging stations at one location replaced the distributed 29 V1G charging stations used in Test~1. Since the PowerFlex interface did not permit actuating individual stations, the 17 charging stations participated in the test as a single aggregate DER. The 0930 – 1010 PT timing of the V1G EV part of the test coincided with the start of the workday and a V1G EV population that had only recently plugged in and therefore had ample remaining charging capacity. The EVs were contributed by UCSD employees and visitors randomly plugging in at the PowerFlex charging stations just before the start of the trial. An aggregate signal of 15~kW to 19~kW was distributed equally amongst the 17 EVs.

\revision{In addition to the new V1G EVs, the V2G population in Test~2 was replaced with a different set of Nuvve chargers to resolve a tracking/noise issue during discharge-to-grid observed in Test~1 and expanded to include an additional charger.}

\subsubsection{Computing Setup} \revision{Test~2 featured a fully distributed architecture 
that consisted of a network of Raspberry Pi’s that asynchronously communicated with each other via an ethernet switch.} In addition, a modified synchronization technique was implemented in the software which improved the fidelity and robustness of message-passing. This upgraded message-passing framework and synchronization technique for both software and hardware resulted in significantly faster communication between nodes.
\subsubsection{\revision{Actuation}} \revision{The order of AHU actuation was modified in Test~2 to allow for device settling time and prevent interference. In particular, in Tests~0 and 1, individual AHUs were ordered and actuated using a protocol that was not cognizant of settling times or building groupings, while the protocol was revised in Test~2 to systematically command the entire population of AHUs in a manner which maximized time between consecutive actuations for an individual unit.}

Test~2 also featured a two-stage approach of actuation that was a result of the DER tracking behavior in Test~1. Some DERs, such as BESS, V1G EVs and V2G EVs, tracked quickly and accurately, whereas others, such as AHUs, tracked poorly. The overall tracking performance in Test~2 was improved by using ``well-behaved" DERs to compensate for AHU tracking errors by incorporating the error signal from actuating AHUs in Stage 1 to the cumulative target signal for BESS, V1G EVs and V2G EVs in Stage 2. Although synchronous actuation of all participating DERs is preferred in practice, the two-stage approach highlights the significance of systematic characterization of DERs in  minimizing~ACE.

\section{\revision{Test Results}} \label{sec:results}
\subsection{Distributed Optimization/Cyber-Layer Results}

In Table~\ref{table:error}, we present MSE results of our 1~s real-time Raspberry-pi distributed optimization solutions (the ``cyber-layer'' of the system). 

\begin{table}[htb]
\centering
\caption{Normalized mean-squared-error of distributed solutions obtained from real-time 1~s intervals compared to centralized solver solution for Test~2 (Section \ref{error-metrics})}\label{table:error}
\begin{tabular}{|c|c|c|c|c|}
\hline
\textbf{DER Type}  & \textbf{RC} & \textbf{PD} & \textbf{DANA} & \textbf{all} \\ \hline
AHU & $0$ & $1.4\times 10^{-7}$ & $2.8\times 10^{-9}$ & $4.6\times 10^{-8}$ \\ \hline
V1G EVs & $0$ & $7.0 \times 10^{-8}$ & $1.7\times 10^{-9}$ & $2.3\times 10^{-8}$ \\ \hline
V2G EVs & $0$ & $6.6\times 10^{-5}$ & $5.0\times 10^{-7}$ & $2.1\times 10^{-5}$ \\ \hline
BESS & $0$ & $2.0\times 10^{-6}$ & $9.1\times 10^{-8}$ & $6.5\times 10^{-7}$ \\ \hline
Total & $0$ & $1.8\times 10^{-5}$ & $1.1\times 10^{-7}$ & $4.9\times 10^{-6}$ \\ \hline
\end{tabular}
\end{table}

RC converged to the exact solution in all instances. This is unsurprising, as the RC problem formulation does not account for individual DER costs and thus, is a much simpler problem with a closed-form solution. For PD and DANA, we obtained excellent convergence, with errors on the order of $0.001\%$ in the worst cases. In general, DANA tended to converge faster than PD \revision{in the sense that the obtained solutions were more accurate under the same fixed 1~s computation time}. For our application with 1~s real-time windows, accuracy and convergence differences did not affect the physical layer results in any tangible way, but applications with more stringent accuracy or speed requirements may benefit from using a faster algorithm like DANA. The differences between DER populations can be largely attributed to the faster time scale of the V2G EVs (and to a lesser extent the BESS), see Table~\ref{table:devices}. Since the V2G EVs were responsible for the high-frequency component of $\Pref$, the solver was required to converge to new solutions at every time step, which induced more error compared to the slow V1G EVs and AHUs with relatively static solutions.

\subsection{Physical-Layer Test Results}\label{ssec:phys-results}
\begin{figure}[hbt!]
\centering 
\includegraphics[width=\linewidth]{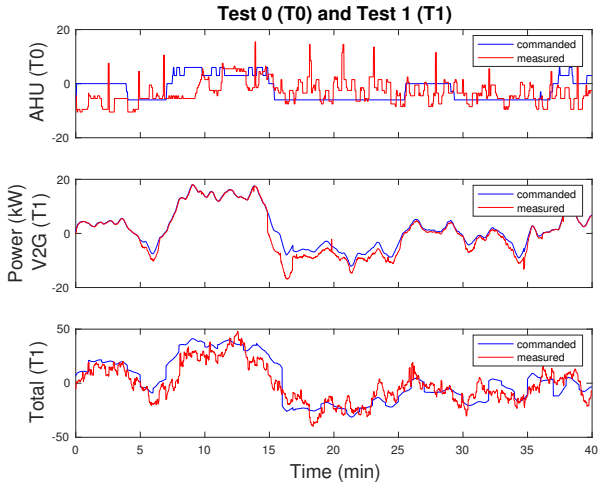}
\caption{\revision{Selected plots from Tests~0 and~1.} \textbf{Top:} AHU response in Test~0. \revision{Note the poor tracking and spikes in the measured response.} \textbf{Middle:} V2G response in Test~1. \revision{Note the inaccuracy in tracking during discharge-to-grid phases.} \textbf{Bottom:} Total response in Test~1. \revision{Note the large-magnitude, low-frequency features demonstrating some broad tracking behavior, but overall poor performance.} }\label{fig:trial0trial1}
\vspace*{-2ex}
\end{figure}

\begin{figure}[hbt!]
\centering 
\includegraphics[width=\linewidth]{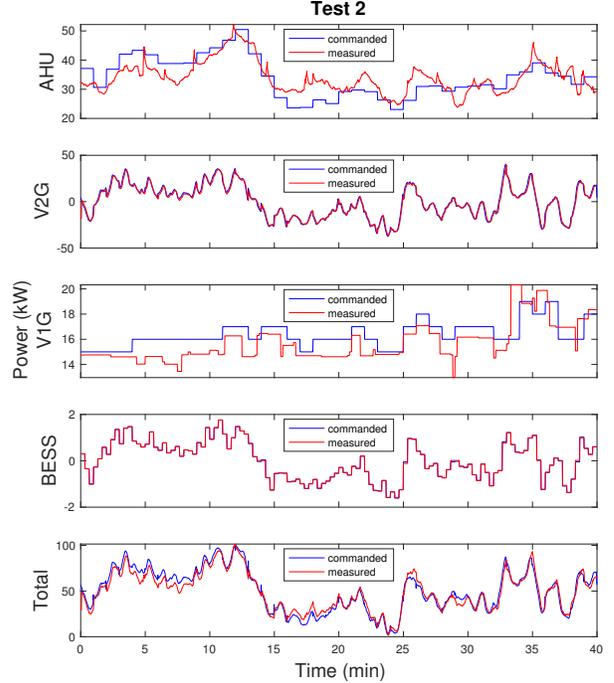}
\caption{\revision{Test~2 results.} From \textbf{top to bottom}, AHU, V2G EVs, V1G EVs, BESS, and total responses. \revision{Note the substantially improved AHU, V2G, and total tracking performance compared to Figure~\ref{fig:trial0trial1}. }}\label{fig:trial2}
\vspace*{-2ex}
\end{figure}

We now present the results of the tracking performance pertaining to the physical-layer of the experiment. We provide only some selective plots for Test~0 and Test~1 in Fig.~\ref{fig:trial0trial1}, and a complete set of plots for each Test~2 DER population in Fig.~\ref{fig:trial2}. Error and tracking delay data defined in Section \ref{error-metrics} is given in Table~\ref{table:phys-error} for Test~1 and Test~2. Data for Test~0 is omitted due to its preliminary nature. The optimal shift described in Section \ref{error-metrics} is applied to each time series and hence some areas in plots may appear like the provided signal anticipated the target.

Signal tracking accuracy in Test~0 was generally poor despite the small number of DERs employed, largely due to inexperience in actuating the AHUs and V1Gs. In particular, Fig.~\ref{fig:trial0trial1} reveals some oscillations in the AHU response. It is overall difficult to determine if even large-feature, low-frequency components of the signal were tracked. Further, data gathering for V1Gs and AHUs was done via noisy and unreliable building ION meters, which motivated the need for outlier treatment (Section~\ref{power-measurements}) in Tests~1 and~2, and resulted in the smoother and better tracking signal in the top plot of Fig.~\ref{fig:trial2}. 

Test~1 yielded a 
111\% RMSE
for AHUs. We speculate that the small 4~s delay \revision{in Test~1} is not representative of the actual AHU delay due to random correlations dominating the time shift for this large error. This is confirmed by a much better AHU response in Test~2 with RMSE 
12\%, where a 105~s delay is more likely to be representative of the true AHU actuation delay.
Given the poor visibility into AHU and V1G controllers explained in Section~\ref{sec:test-scenario}, it is challenging to identify the source of the poor tracking behavior. We speculate that DER metering at the building level rather than the DER level was a major source of error for AHU and V1G in Test~1. This was largely resolved in Test~2 by utilizing a different population of V1Gs with dedicated meters and by modifying the actuation scheme for AHUs to be less susceptible to metering errors as described in Section~\ref{test-2}. Additionally, the actuation-interface stalling for V1G EVs, described in Section~\ref{test-1}, was dominant in Test~1, resulting in the poor tracking for V1Gs. Actuating-interface issues were resolved in Test~2 by utilizing an automated control scheme for the V1Gs, which led to significantly lower error.

The BESS emerged as the star performer achieving very accurate tracking across all tests with no delay. The V2G EVs also performed relatively well aside from a signal overshoot issue observed during the discharge cycle in Test~1 seen in Fig.~\ref{fig:trial0trial1}. The issue was resolved in Test~2 by using V2G EV charging stations from a different manufacturer (Princeton Power), as described in Section \ref{test-2}. The V2G charging stations deployed for these tests were pre-commercial or early commercial models that had a few operating issues, such as the overshoot issue during Test~1.

The inability of the AHUs to respond to steep, short ramps (Fig.~\ref{fig:trial2}) could be due to slow start-up sequences programmed into the building automation controllers to increase device longevity or due to transients associated with driving their AC induction electric motors. Tackling this would require dynamic models and parameter identification of signal response and delay. With the new V1G EV population in Test~2, tracking delay reduced from 40~s to 10~s and the tracking accuracy improved significantly. The 1~kW bias seen in Fig.~\ref{fig:trial2} is likely due to rounding errors arising from the inability of PowerFlex charging stations to accept non-integer setpoints.

The superior performance of the BESS and V2Gs motivated the two-stage actuation scheme described in Section~\ref{test-2}, which contributed to reducing the total RMSE from 
50\% in Test~1 to 
10\% in Test~2 (compare the bottom plots of Figs~\ref{fig:trial0trial1} and~\ref{fig:trial2}). The two-stage approach allows a sufficiently large proportion of accurately tracking DERs to compensate for the errors of the first stage, where tracking is worse. In this way, poorly-tracking DERs, such as AHUs, can still contribute by loosely tracking some large-feature, low-frequency components of the target signal. The low-frequency contribution reduces the required total capacity of the strongly-performing DERs in the second stage leading to more fine-tuned signal tracking in aggregation. Some recommended rules of thumb for two-stage approach are: (i) Total capacity of first-stage DERs is less than or equal to total capacity of second-stage DERs. (ii) DERs in the first stage are capable of tracking with $<$ 50\% RMSE. (iii) DER cost functions are such that the deviation from the baseline is lower cost for first-stage DERs than for second-stage. (iii) allocates a significant portion of the target signal initially to first-stage DERs, freeing up DER capacity in the second-stage for error compensation. 

\begin{table}[htb]
\caption{\textbf{Left:} Relative root mean-squared-error of tracking error by DER type. \textbf{Right:} Delay (optimal time-shift) of DER responses in seconds.}\label{table:phys-error}
\centering
\begin{tabular}{|c|c|c|}
\hline
\textbf{DER Type}  & \textbf{Test~1} & \textbf{Test~2} \\ \hline
AHU & 
1.11 & 
0.12\\ \hline
V1G EVs & 
0.68 & 0.077 \\ \hline
V2G EVs & 
0.30 & 
0.060 \\ \hline
BESS & 0.054 & 0.018 \\ \hline
Total & 
0.50 & 0.097 \\ \hline
\end{tabular}
\qquad
\begin{tabular}{|c|c|c|}
\hline
\textbf{DER Type}  & \textbf{Test~1} & \textbf{Test~2} \\ \hline
AHU & 4 & 105 \\ \hline
V1G EVs & 40 & 10 \\ \hline
V2G EVs & 5 & 3 \\ \hline
BESS & 0 & 0 \\ \hline
Total & N/A & N/A \\ \hline
\end{tabular}
\end{table}

\subsection{Economic Benefit Analysis}

Here, we evaluate the economic benefit of the proposed test system, which is vital for wider scale adoption of DERs as a frequency regulation resource in real electricity markets. To this end, we take an approach similar to~\cite{YL-PB-SM-TM:15} to first demonstrate that the testbed is eligible to participate in the PJM ancillary services market. Following the PJM Manual 12~\cite{PJM:20} (Section~\ref{error-metrics}), we compute a Correlation Score $S_c$ = 
0.98, Delay  Score $S_d$ = 
0.65, and Precision Score $S_p$ = 
0.91 from data for Test~2, and obtain a Performance Score $S = 
0.85\geq 0.75$, which confirms the eligibility to participate in the PJM ancillary service market.

Next, we compute the estimated annual revenue assuming that the resources are available throughout the day. \revision{Using PJM's ancillary service market data\footnote{\url{https://dataminer2.pjm.com/feed/reg_prices/definition}} 
with our total (active) DER capacity of 184~kW and performance score of 
0.85, the capability and performance credits for this population of resources (cf.~\cite[Section 4]{PJM:19}) would respectively be \$135 and \$11, for July 9, 2020. This gives an estimated amount of \$53,290 as the total annual revenue.} Note that the 184~kW DER capacity employed in this work represents less than 5\% of the total DER capacity and less than 0.5\% of the total capacity of the UCSD microgrid, cf.~\cite{BW-JD-DW-JK-NB-WT-CR:13}. As such, the revenue would significantly increase if more microgrid resources are utilized for regulation, even with reduced availability.

\section{Conclusions}
We have presented one of the first real-world demonstrations of secondary frequency response in a distribution grid using up to 176 heterogeneous DERs. The DERs include AHUs, V1G and V2G EVs, a BESS, and passive building loads and PV generators. The computation setup utilizes state-of-the-art distributed algorithms to find the solution of a power allocation problem. We show that the real-time distributed solutions are close to the true centralized solution in an MSE sense. Tests with real, controllable DERs at power closely track the given active-power reference signal in aggregation. \revision{Further, our economic benefit analysis shows a potential annual revenue of \revision{\$53K} for the chosen DER population. These tests highlight the importance of dedicated and noise-free measurement sensors and a well-understood and reliable DER control interface for precise signal tracking. } \revision{Extensions of this work are ongoing under DERConnect\footnote{\url{https://sites.google.com/ucsd.edu/derconnect/home}}, a new project at UCSD that aims to develop a testbed consisting of 2500 DERs  that allows for online implementation of various distributed algorithms.}
As is already recognized by the power systems community and  federal funding agencies such as ARPA-e and National Science Foundation, large-scale power-in-the-loop testing is needed for transitioning distributed technologies to real distribution systems. 
We hope that this work spurs further testing and ultimately widespread adoption of coordinated resource control algorithms by relevant players in industry.

 \appendices

 \section{Distributed Coordination Algorithms}\label{sec:appendix}

 In this section we describe the algorithms used in our distributed computing platform to solve~\eqref{eq:opt}. 

 \emph{Ratio-Consensus (RC)}: The ratio-consensus of~\cite{ADDG-CNH-NHV:12} computes equitable contributions from all DERs without DER-specific cost functions (or constant DER costs). The ratio-consensus algorithm for providing $\Pref$ is given by
 \begin{alignat*}{2}
     y_i[k+1] &= \sum_{j\in\N_i} \frac{1}{\vert \N_i\vert}y_j[k], &
     z_i[k+1] &= \sum_{j\in\N_i} \frac{1}{\vert \N_i\vert}z_j[k], \\
     y_i[0] &= \begin{cases} \frac{\Pref}{\vert\I\vert} - \pu_i, & i\in\I, \\
     -\pu_i, & i\notin\I,
     \end{cases} & 
     z_i[0] &= \po_i - \pu_i,
 \end{alignat*}
 where, $k$ is the iteration number, $y_i$ and $z_i$ are two auxiliary variables maintained by each agent, $\N_i$ denotes the neighboring DERs of DER $i$, and $\pu_i$ and $\po_i$ are the minimum and maximum power level for DER $i$ from the problem formulation in Section~\ref{optimization-statements}. $\I$ denotes the subset of DERs which know the value of the reference signal. One can see~that 
 \begin{equation*}
 \begin{aligned}
     p_i^\star &= \pu_i + \underset{k\rightarrow\infty}{\lim} y_i[k]/z_i[k](\po_i - \pu_i) \\
     &= \pu_i + \frac{\Pref - \sum_i \pu_i}{\sum_i \po_i - \pu_i}(\po_i - \pu_i),
 \end{aligned}
 \end{equation*}
 where $p_i^\star$ is then the power assignment for DER $i$.

 \emph{Primal-Dual (PD)}: Both this dynamics and DANA (described next) take into account the cost functions of the DER types when computing the power setpoints, i.e., $f_i$ are nonconstant. These functions are modeled as quadratics, which is a common choice in generator dispatch~\cite{AW-BW-GS:12}. The dynamics is based on the discretization of the primal-dual dynamics~\cite{AC-BG-JC:17-sicon} for the augmented Lagrangian of the equivalent reformulated problem, see~\cite{AC-JC:16-allerton}, and it has a linear rate of convergence to the optimizer. The algorithm is given by
 \begin{equation*}
     \begin{aligned}
         \begin{bmatrix}
             \dot{p}_i \\ \dot{y}_i \\ \dot{\lambda}_i
         \end{bmatrix} =
         \begin{bmatrix}
         -\left(f_i'(p_i)+ \lambda_i + p_i \sum_{j\in\N_i} L_{ij}y_j - \Pref/n \right) \\
         -\left( \sum_{j\in\N_i} L_{ij} (\lambda_j + x_j - \Pref/n) + \sum_{j\in\N_i^2} L_{ij}^2 y_j  \right) \\ 
         p_i + \sum_{j\in\N_i} L_{ij} y_j - \Pref/n
         \end{bmatrix},
     \end{aligned}
 \end{equation*}
 where, $L$ is the Laplacian matrix of the communication graph (see~\cite{FB-JC-SM:09}), $y_i$ is an auxiliary variable, and $\lambda_i$ is the dual variable associated with agent $i$. The update step is followed by a projection of the primal variable $p_i$ onto the box constrained local feasible set. These dynamics converge from any set of initial conditions. Since this algorithm evolves in continuous time, we use an Euler discretization with fixed step-size to implement it in discrete time.

 \emph{Distributed Approximate Newton Algorithm (DANA)}: The Distributed Approximate Newton Algorithm (DANA) of~\cite{TA-CYC-SM:18-auto} has an improved rate of convergence compared to PD. This algorithm solves the equivalent reformulated problem
 \begin{equation}\label{eq:DANA-opt}
 \begin{aligned}
 \underset{z\in\real^n}{\text{min}} \
 &f(p^0 + Lz) = \sum_{i=1}^n f_i(p_i^0 + L_i z), \\
 \text{subject to} \ &\pu - p^0 - Lz \leq \zeros_n, \\
 &p^0 + Lz - \po \leq \zeros_n,
 \end{aligned}
 \end{equation}
 where $p^0$ is a vector of initial power levels of all the DERs with $\sum_i p_i^0 = \Pref$, and $z$ is the new variable of optimization. The continuous time dynamics are given by
 \begin{equation*}
 \begin{aligned}
     \dot{z} &= -A_q\nabla_z \Lagr (z,\lambda), \\
     \dot{\lambda} &= [\nabla_\lambda \Lagr (z,\lambda)]^+_\lambda ,
 \end{aligned}
 \end{equation*}
 where $\Lagr$ is the Lagrangian of~\eqref{eq:DANA-opt} and $A_q$ is a positive definite weighting on the gradient direction which provides distributed second-order information. For brevity, we do not provide the full details of the algorithm here, which can instead be found in~\cite{TA-CYC-SM:18-auto}. The cost functions are again taken to be quadratic with strictly positive leading coefficients.

\section*{Acknowledgements}
We would like to thank numerous people in the UCSD community and beyond for their generous contributions of time and resources to enable such an ambitious project to come together. We extend thanks to: (i) Aaron Ma and Jia (Jimmy) Qiu for assisting with hardware setup and software development for the distributed computation systems; (ii) Kevin Norris for coordinating the fleet vehicles; (iii) Abdulkarim Alamad for overseeing V1G drivers in Test~2; (iv) Kelsey Johnson for managing the Nuvve contributions; (v) Ted Lee, Patrick Kelly, and Steven Low for managing the PowerFlex contribution; (vi) Marco Arciniega, Martin Greenawalt, James Gunn, Josh Kavanagh, Jennifer Rodgers, Patricia Roman and Lashon Smith from UCSD parking
for reserving EV charging station parking spaces; (vii) Charles Bryant, Harley Crace, John Denhart, Nirav Desai, John Dilliott, Mark Gaus, Martin Greenawalt, Gerald Hernandez, Brandon Hirsch, Mark Jurgens, Josh Kavanagh, Jose Moret, Chuck Morgan, Curt Lutz, Jose Moret, Cynthia Wade, Raymond Wampler and Ed Webb for contributing their EVs in Test~1; (viii) Adrian Armenta, Adrian Gutierrez and Minghua Ong who helped with ChargePoint manual control; (ix) Bob Caldwell (Centaurus Prime), Gregory Collins, Charles Bryant, and Robert Austin for programming and enabling the AHU control; (x) Gary Matthews and John Dilliott for permitting the experimentation on ``live'' buildings and vehicles; and (xi) Antoni Tong and Cristian Cortes-Aguirre for supplying the BESS. Finally, we would like to extend a sincere thanks to the ARPA-e NODES program for its financial support and to its leadership, including Sonja Glavaski, Mario Garcia-Sanz, and Mirjana Marden, for their vision and push for the development of large-scale power-in-the-loop testing environments.

\bibliographystyle{IEEEtran}
\bibliography{alias,SMD-add,JC,SM,Main-add}

\end{document}